\documentclass{elsart}
\usepackage{graphicx, epsfig}
\usepackage{amssymb}
\usepackage{amsmath}
\usepackage{bm}
\begin{document}

\begin{frontmatter}
\title{Velocity Selection in 3D Dendrites: \\Phase Field Computations and Microgravity Experiments}
\author[a]{Y. B. Altundas},
\ead{baltundas@slb.com}
\author[b]{G. Caginalp}
\ead{caginalp@pitt.edu}
\ead[url]{http://www.pitt.edu/$\sim$caginalp}
\address[a]{Institute for Mathematics and its Applications, University of Minnesota\\Minneapolis, MN 55455}
\address[b]{Department of Mathematics, University of Pittsburgh\\ Pittsburgh, PA 15260}

\begin{abstract}
The growth of a single needle of succinonitrile (SCN) is studied
in three dimensional space by using a phase field model. For
realistic physical parameters, namely, the large differences in
the length scales, i.e., the capillarity length ($10^{-8}{
cm}$~-~$10^{-6}{ cm}$), the radius of the curvature at the tip of
the interface (${10^{-3}}{ cm}$~-~${10^{-2}}{~cm}$) and the
diffusion length (${10^{-3}}{ cm}$~-~$10^{-1}{ cm}$), resolution
of the large differences in length scale necessitates a $500^{3}$
grid on the supercomputer. The parameters, initial and boundary
conditions used are identical to those of the microgravity
experiments of Glicksman \textit{et al} for SCN. The numerical
results for the tip velocity are (i) largely consistent with the
Space Shuttle experiments; (ii) compatible with the experimental
conclusion that tip velocity does not increase with increased
anisotropy; (iii) different for 2D versus 3D by a factor of
approximately $1.9$; (iv) essentially identical for fully versus
rotationally symmetric 3D.
\end{abstract}
\end{frontmatter}
\maketitle
\section{Introduction}
The temporal evolution of an interface during solidification has
been under intensive study by physicists and material scientists
for several decades. The interface velocity and shape have
important consequences for practical metallurgy, as well as the
theory, e.g., velocity selection mechanism and nonlinear theory of
interfaces.

The simplest observed microstructure is the single needle crystal
or dendrite, which is observed to be a shape resembling a
paraboloid (but not fully rotationally invariant away from the
tip) growing at a constant velocity, $v_{0}$, with tip radius,
$R_{0}$.

An early model of this phenomenon by Ivantsov~\cite{Ivantsov1947}
stipulated the heat diffusion equation in one of the phases and
imposed latent heat considerations at the interface. With the
interface assumed to be at the melting temperature, the absence of
an additional length scale implies the existence of an infinite
spectrum of pairs of velocities and tip radii, $(v_{0},R_{0})$.
Since the experimental results have shown that there is a unique
pair $(v_{0},R_{0})$ that is independent of initial conditions,
there has been considerable activity toward uncovering the
theoretical mechanism for this velocity selection (see, for
example~\cite{BenJacob84,KessKopLev1984,BrenerMelnikov1991,PomeauAmar1991}).
The emergence of the capillarity length associated with the
surface tension as an additional length scale has provided an
explanation for the selection mechanism. Advances in computational
power and a better understanding of interface models and their
computation have opened up the possibility of comparing
experimental values for the tip velocity with the numerical
computations. This is nevertheless a difficult computational issue
in part due to the large differences in length scales that range
from $1~cm$ for the size of the experimental region, to $14$
microns for the radius of curvature near the tip, $10^{-6}{~cm}$
for the capillarity length, to $10^{-8}{~cm}$ interface thickness
length.

One perspective into the theoretical and numerical study of such
interfaces has been provided by the phase field model introduced
in~\cite{CaginalpCM82,CaginalpLN84} in which a phase, or order
parameter, $\varphi$, and temperature, $T$, are coupled through a
pair of partial differential equations described below (see also
more recent
papers~\cite{Almgren1999,KimProvGoldDant1999,HariharanYoung2001}).
In physical terms, the width of the transition region exhibited by
$\varphi$ is Angstroms. In the $1980$'s three key results
facilitated the use of these equations for computation of physical
phenomena. If the equations are properly scaled one can (i)
identify each of the physical parameters, such as the surface
tension, (ii) and attain the sharp interface problem as a
limit~\cite{CaginalpArchive1986}, and (iii) use the interface
thickness, $\varepsilon ,$ as a free parameter, since the motion
of the interface is independent of this parameter
~\cite{CaginalpSocolov1989}. This last result thereby opened the
door to computations with realistic material parameters, by
removing the issue of small interface thickness. However, the
difference in scale between the radius of the curvature and
overall dimensions still pose a computational challenge. More
recently, several computations, have been done using the phase
field
model~\cite{AbBrKr1997,KarRap1998,WarnerKoCa2000,RegWilPoLi2001,ProvGDCLKGA1999,WangSWMCBMcF1993},
with some 3D computations in~\cite{KarRap1998} utilizing the model
and asymptotics of~\cite{CaginalpCh1991}, that will be compared
with our results below. Also, George~\textit{et al} studied the
simulation of dendritic growth in three dimensional space using a
phase field model~\cite{GeorgeWarren2002}.

Our work differs from the works referenced above in many aspects.
However, the main difference arises from the adaptation of the
experimental conditions in the simulation of dendritic growth.
Most importantly, we use true values of physical parameters which
are obtained from the microgravity experiment for
SCN~\cite{KossGlickBuCoWi1995}. In order to deal with different
length scales and the diffusion during freezing in the thin
interfacial region, we implement a fully parallel architecture in
a three dimensional space which enables us to use enough grid
points and perform an efficient calculation.

Solidification is a complicated nonlinear process. Modeling
necessarily involves making choices of physical effects that are
to be included in the equations. Comparison of computations with
experiments that are closest to the mathematical model yields the
most convincing test of the model and computations. The modeling
of single-needle dendrites has usually been carried out using the
diffusion equation as a mechanism for the dissipation of heat.
However, all of the experiments until the Space Shuttle
experiments had been done under conditions of normal gravity, so
that convection in the liquid is an important mechanism for the
dissipation of the latent heat released at the interface. The
microgravity experiments performed on the Space
Shuttle~\cite{KossGlickBuCoWi1995} provide the first opportunity
to test whether the mathematical models agree with experiments,
since the absence of gravity essentially eliminates convection,
thereby leaving diffusion as the main mechanism for heat transport
away from the interface.

While numerous computer calculations have been performed on both
sharp interface and phase field models of solidification,
comparison with experiment has always been a difficulty due to the
vastly different length scales in the problem (e.g 10$^{-6}{~cm}$
for capillarity length and $1~cm$ for the overall dimensions of
the experiment), and the three dimensional nature of the problem.
In the absence of direct comparison with experiment, it is also
difficult to know whether some of the simplifications that have
been used, such as setting the kinetic coefficient, $\alpha$, to
zero are valid.

In this paper we perform large scale 3D parallel computations of a
phase field model with the modification introduced in
~\cite{CaginalpCh1991}. The key aspects of these computations are
summarized below.

(A) We perform fully three dimensional parallel computations by
adopting the experimental conditions used in the Space Shuttle
experiment. The symmetry is utilized only along the major axes
(rather than rotational symmetry). This allows us to compare the
tip velocity with the actual experiments in a meaningful way. The
calculations utilize the parameters and boundary conditions of the
IDGE microgravity experiments for
SCN~\cite{KossGlickBuCoWi1995,GlickSchAy1977}. All previous
experiments done under normal gravity conditions introduced
convection. Hence this provides an opportunity to compare
experiments in the absence of convection to theory that also
excludes convection. The difference between the experimental
results and our computations thereby defines the challenges for
additional physical effects that need to be modeled.

(B) The role of anisotropy in velocity selection has been noted in
the computational references cited above. Glicksman and
Singh~\cite{GlickSingh1989} compare experimental tip velocity of
SCN with pivalic acid (PVA) whose coefficient of surface tension
anisotropy (defined below) differs by a factor of $10$ but are
otherwise similar, except perhaps for the kinetic coefficient. We
perform two sets of calculations in which all parameters are
identical (SCN values) except for the anisotropy coefficient. Our
computations confirm (consistent with the experimental
results~\cite{KossGlickBuCoWi1995}) that the velocity is nearly
identical when the magnitude of the anisotropy is varied by a
factor of $10$ with all other parameters fixed (at the SCN
values).

(C) Most of the previous numerical computations that simulate the
interface growth were done in two dimensional space. Our
computations shows that the 2D and 3D computations differ by a
factor of approximately $1.9$. The results of the 3D for tip
velocity can also be compared with our previous
computations~\cite{AltunCag} that utilized rotational symmetry to
reduce the 3D computations to two computational spatial
dimensions.

(D) The role of the kinetic coefficient [see definition of $\alpha
$ below equation (\ref{Phase1})] is subtle, and this material
parameter is often set to zero, for convenience, in theoretical
and computational studies. We find, however, that there is a
significant difference in the tip velocity when all other
parameters are held fixed while this coefficient is varied.
Consequently, this kinetic coefficient may be of crucial
importance in determining the selection of tip velocity. A better
understanding of this issue may lead to theory that can explain a
broader range of undercooling and velocity.
\section{Mathematical Modeling}
In the computations below, we use a version of the phase field
equations introduced in~\cite{CaginalpCh1991}, for which the phase
or order parameter, $\varphi(\vec{x},t)$, as a function of spacial
point, $\vec{x}$, and time, $t$, is exactly $-1$ in the solid and
$+1$ in the liquid. The order parameter is coupled with the
dimensionless temperature, $u$, which is given by the following
relation along with the capillary length, $d_0$.
\begin{equation}\label{dimlesstemp}
u(x,t)=\frac{T-T_m}{l_v/c_v},~~d_0=\frac{\sigma c_v}{[s]_{E} l_v}
\end{equation}
where $T_m$,~$l_v$,~$c_v$,~$\sigma$ and $[s]_E$ are the melting
temperature, latent heat, specific heat per unit volume of the
material, surface tension and the difference in the entropy (in
equilibrium) per unit volume between the solid phase and liquid
phase, respectively. Thus, we can define the interface by
$\Gamma=\{x\in\Omega:\varphi(x,t)=0\}$ and write the dimensionless
phase field equations as follows
\begin{equation}\label{Phase1}
\alpha\varepsilon^{2}\varphi_t=\varepsilon^{2}\Delta{\varphi}+g(\varphi)+
\frac{5}{8}\frac{\varepsilon}{d_0}{u}{f'(\varphi)}
\end{equation}
\begin{equation}\label{Phase2}
u_{t}+\frac{1}{2}\varphi_{t}=D\Delta{u}
\end{equation}
where
\begin{equation}
g(\varphi)=\frac{\varphi-\varphi^{3}}{2},~~
f'(\varphi)=(1-\varphi^{2})^{2},~~D=\frac{K}{c_v }
\end{equation}

Here, $\alpha$ is the kinetic coefficient and $\varepsilon$ is the
interface thickness that can be used as a "free
parameter"~\cite{CaginalpSocolov1989}. In the limit as
$\varepsilon$ vanishes as all other parameters held fixed,
solutions to~(\ref{Phase1}) and (\ref{Phase2}) are governed by the
sharp interface model
\begin{equation}\label{Sharp1}
u_t=\nabla\cdot D\nabla u
\end{equation}
\begin{equation}\label{Sharp2}
v_n=-D\left[ {\nabla u\cdot \hat{n}}\right] _{-}^{+}
\end{equation}
\begin{equation}\label{Sharp3}
u=-d_{0}(\kappa+\alpha  v_{n})
\end{equation}

where the parameters $d_0$, $D$ and $\alpha$ are the same as in
the phase field model, and $v_n$ is the interface normal growth
velocity~(with normal $\hat{n}$ chosen from solid ($-$) to liquid
($+$) )~\cite{CaginalpLN84,CaginalpCh1991}.
\subsection{Initial and boundary conditions}
In order to simulate interface growth of a dendrite in 3D, we
choose a cube of ${[-1,1]}^{3}$ which is assumed to be filled with
pure SCN melt initially. The solidification of the melt is
initiated by a small solid SCN ball of radius, $R_0$, which is
placed at the center of the chamber. The temperature at the
boundary is kept at constant undercooling value, $u_{\infty}$, and
the liquid temperature inside the chamber declines exponentially
from $u=u_{solid}$ on the interface of the seed to the boundary of
the chamber. In particular, the initial conditions of $u$ inside
the chamber are given by a plane wave solution
to~(\ref{Sharp1})~and~(\ref{Sharp2}) which is given by
\begin{equation}\label{InitialU}
u_{trav}(z,t)=\left\{\begin{array}{cl}
    u_{\infty}[1-e^{-v(z-vt/|u_{\infty}|)/(D|u_{\infty}|)}], & z\geq vt/|u_{\infty}| \\
    0, & z<vt/|u_{\infty}|
    \end{array}\right.
\end{equation}

where $z$ is the signed distance from the seed interface~(positive
in the liquid) and $u_{\infty}=\frac{T_{\infty}-T_m}{l_v/c_v}$
denotes the dimensionless undercooling value where~$T_{\infty}$ is
the far field temperature. The initial value of $\varphi$ is
obtained from a leading term asymptotic expansion
solution~\cite{CaginalpCh1991}
\begin{equation}\label{InitialF}
\varphi(x,t)=\tanh{\left(\frac{z-vt}{2\varepsilon}\right)}+\textit{higher
order terms}
\end{equation}

\subsection{Implementation of anisotropy}
Anisotropy is important in determining the shape of dendrites that
grow exclusively in the preferred directions. The experimental
evidence shows conclusively that surface tension, $\sigma$,
exhibits anisotropy~\cite{GlickSingh1989}. While there is the
possibility of dynamical (i.e., through $\alpha$ in equation
~(\ref{Phase1}) or (\ref{Phase2})) or other anisotropy the
experimental measurements of anisotropy in these experiments are
confined to those related to surface tension. Surface tension
anisotropy has been modeled in several ways.
\begin{figure}
\begin{center}
\includegraphics*[width=2.5in]{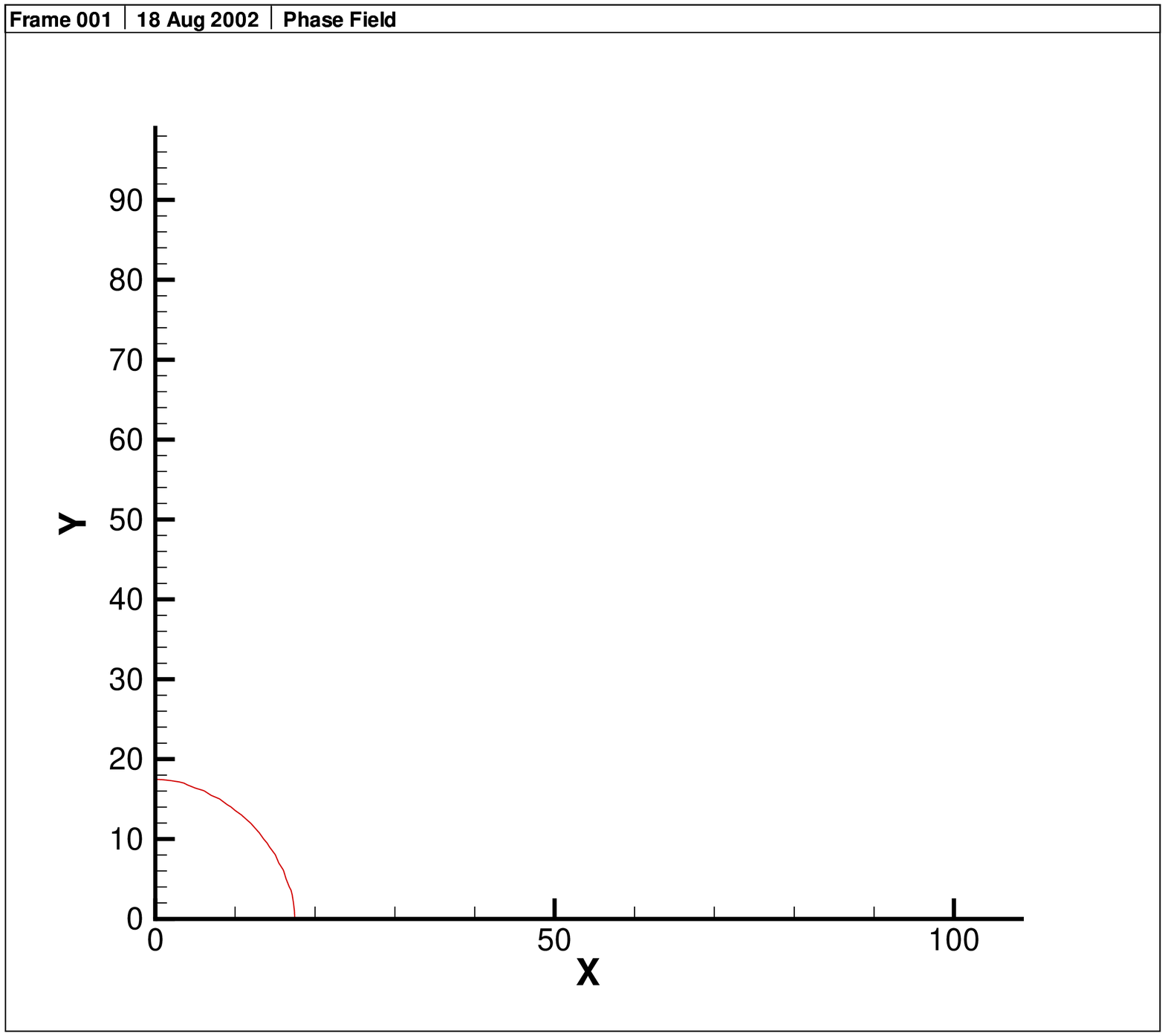}
\includegraphics*[width=2.5in]{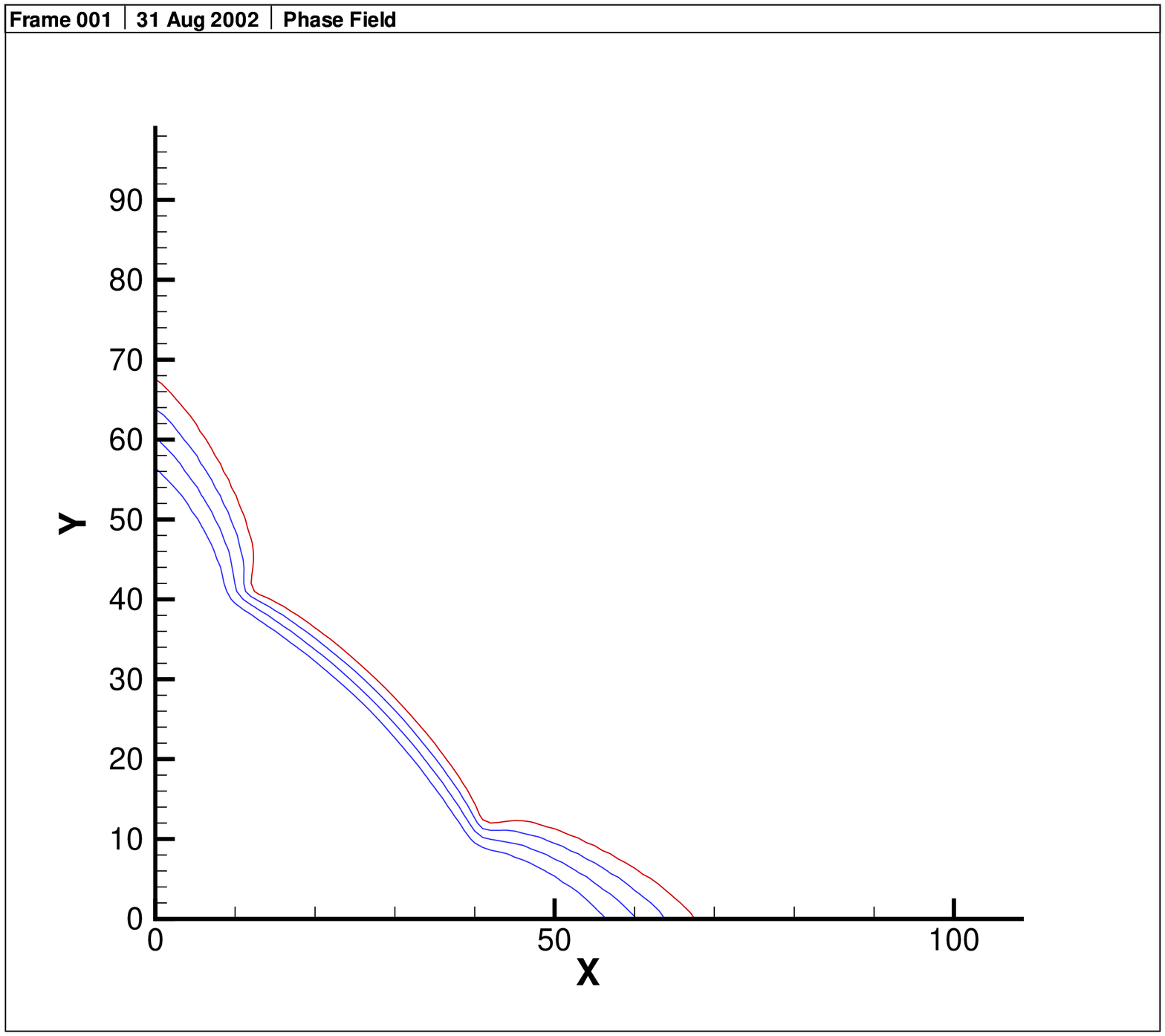}
\end{center}
\caption{\textit{Contour plots of the interface at different times
on xy plane for $u_{\infty}=0.0265$ which shows the effect of the
anisotropy: (a) The position of the interface in 10 seconds; (b)
The position of the interface at latter times (135 sec, 140 sec,
145 sec and 150 sec).}} \label{Fig:ContourPlots}
\end{figure}\bigbreak

Let $\hat{n}=(n_x, n_y, n_z)$ be the normal to the interface. We
utilize the simplest possible function describing the dependence
of surface energy on $\hat{n}$ in the case of an underlying cubic
symmetry. The relation can be given by~\cite{KesLev,WheelMcFad}
\begin{equation}\label{SurfaceTensionN}
\sigma(\hat{n})=a_s
[1+{\epsilon}^{\prime}(n_{x}^{4}+n_{y}^{4}+n_{z}^{4})]
\end{equation}
which is rewritten in terms of spherical angles as
\begin{equation}\label{SurfaceTensionAngles}
\gamma (\theta ,\phi)=a_{s}\{1+\zeta[\cos^{4}{\phi
}+\sin^{4}{\phi}(1-2\sin^{2}{\theta}\cos^{2}{\theta})]\}
\end{equation}
where $\theta$ and $\phi$ are the angles which correspond to the
normal, $n$, with respect to a crystal axis. The parameters,
$\zeta$ and $a_s$, can be related to usual measure of anisotropy
strength, $\delta_{\sigma}$, by the relations
$a_s=(1-3{\delta_{\sigma}})$ and
$\zeta=\frac{4\delta_{\sigma}}{a_s}$~\cite{KarRap1998}.

Aysmptotic analysis~\cite{CaginalpAnPhy} shows that with the
anisotropy, the Gibbs\--Thomson relation~(\ref{Sharp3}) is
modified not only in terms of the angles, but with their second
derivatives. In our simulation, we assume that the dendrites grow
along an axis of symmetry and set $\kappa$ to be the mean
curvature. Thus, we can rewrite the Gibbs\--Thomson
equation~(\ref{Sharp3}) as
\begin{equation}\label{Sharp3_3D_Anis}
u=-~\overline{d}_{0}(\hat{n})\kappa -
\overline{d}_{0}(\hat{n}){\alpha}v_{n}
\end{equation}
where
\begin{equation}\label{Anis_d_0}
\overline{d}_{0}(\theta,\phi)=d_{0}[\gamma
(\theta,\phi)+\frac{\partial^{2}\gamma(\theta,\phi)}{\partial{\theta}^2}+
\frac{\partial^{2}\gamma(\theta,\phi)}{\partial{\phi}^2}]
\end{equation}
\section{Discretization}
A large number of mesh points are necessary in order to calculate
the tip growth velocity and tip radius accurately. The values of
$u$ and $\varphi$ across the interface vary from maximum to
minimum within a short distance. This requires finer grid spacing
so that number of mesh points is adequate to resolve the
interfacial area. We make a physically reasonable but
computationally very useful assumption that the dendrite grows
symmetrically along the coordinate axes. Thus, the computational
domain is reduced to $\Omega=[0,1]^{3}$ which decreases the
overall grid usage by $7/8$. In this work, we use $400\--500$
uniform grid points on each side of $\Omega$ which guarantees that
we have at least $6$\--$7$ grid points located on the interface
region as measured from $\varphi=-0.9$ to $\varphi=0.9$.

In order to discretize the equations~(\ref{Phase1})
and~(\ref{Phase2}), let $\varphi^{p}_{ijk}$ and $u^{p}_{ijk}$
denote the discrete values of $\varphi(x_i,y_j,z_k,t_p)$ and
$u(x_i,y_j,z_k,t_p)$, respectively. We lag the nonlinear terms
in~(\ref{Phase1}) and discretize the equations~(\ref{Phase1})
and~(\ref{Phase2})~by using semi\--implicit Crank\--Nicolson
finite difference method. Thus, we write the discrete equations as
follows:
\begin{equation}\label{DiscrtePhase1}
\delta_{t}^{+}u_{ijk}^{p}=\frac{D}{2}\{\Delta{u}^{p}_{ijk} +
\Delta{u}^{p+1}_{ijk}\}
-\frac{1}{2}\delta_{t}^{+}\varphi_{ijk}^{p}
\end{equation}
\begin{equation}\label{DiscrtePhase2}
\varepsilon^{2}\alpha\delta_{t}^{+}\varphi_{ijk}^{p}=\frac{\varepsilon^{2}}{2}
\{\Delta{\varphi}^{p}_{ijk} + \Delta{\varphi}^{p+1}_{ijk}\}+
g(\varphi_{ijk}^{p})+
\frac{5}{8}\frac{\varepsilon}{d_{0}}u_{ijk}^{p}f'(\varphi_{ijk}^{p})
\end{equation}
where
\begin{equation}
\Delta{\phi^{T}_{ijk}}=\delta^{\pm}_{x_{i}}\phi_{ijk}^{T}+
\delta^{\pm}_{y_{j}}\phi_{ijk}^{T}+\delta^{\pm}_{z_{k}}\phi_{ijk}^{T}
\end{equation}

for $i, j, k=0,1,2,...,N-1$ and $p=0,1,2,...,T_{final}$. The
operators such as $\delta^{\pm}_x$~and~$\delta_{t}^{+}$ in the
equations~(\ref{DiscrtePhase1})~and~(\ref{DiscrtePhase2}) are the
finite difference operators and they can be given as
\begin{equation}
\delta_{x}^{\pm}\phi(x,y,z,t)=\frac{\phi(x+\Delta{x},y,z,t)
+\phi(x-\Delta{x},y,z,t)-2\phi(x,y,z,t)}{(\Delta{x})^{2}}
\end{equation}
and
\begin{equation}
\delta_{t}^{+}\phi(x,y,z,t)=\frac{\phi(x,y,z,t+\Delta{t})-\phi(x,y,z,t)}{\Delta{t}}\cdot
\end{equation}
\section{Values of SCN Parameters}
The comparison of the computational results with experiments makes
sense only if we use the true value of the physical parameters.
Throughout the simulation, the parameters $d_0$ and $\sigma_{0}$
for SCN are set to be $2.83\times 10^{-7}~cm$ and
$8.9~ergs/cm^{2}$, respectively~\cite{GlickSchAy1977}. The
diffusivity parameters, $D_{liquid}$ and $D_{solid}$, are almost
the same for the liquid and solid SCN. Thus, we set the
diffusivity $D$ to be $1.147\times 10^{-3}cm^{2}/sec$ which is the
value for $D_{liquid}$~\cite{GlickSchAy1977}.

All of the parameters in the phase field equations are physically
measurable quantities, including $\varepsilon$ which is a measure
of the interface thickness. It was shown
in~\cite{CaginalpLN84,CaginalpArchive1986} that the solutions of
the phase field equations (scaled in this form) approach those of
the sharp interface model~(\ref{Sharp1})-(\ref{Sharp3}) provided
all other parameters are held fixed as $\varepsilon$ approaches
zero. The rate of convergence, however, emerges as a key issue. In
particular, the true size of $\varepsilon$ is a few atomic
lengths, or Angstrom, while the size of experimental region is at
least $1cm$. Thus using the true value of $\varepsilon$ would
necessitate $10^{9}$ grid points in each direction yielding
unfeasible computation. From a computational perspective, one
needs to have at least several grid points in the interfacial
region in order to accurately calculate the derivatives of order
parameter that implicitly define the surface tension.

A computational breakthrough was the discovery that $\varepsilon$
could be made many orders of magnitude larger without influencing
the interface motion. Caginalp and
Socolovsky~\cite{CaginalpSocolov1989,CaginalpSocol1991} showed
that so long as one chooses an appropriate number of grid points
in the interface region (defined by the magnitude of $\varepsilon$
and $h$ denotes the uniform grid size), guaranteed by the range
$0.75<\frac{\varepsilon}{h}<1.1$, one can resolve the motion
accurately. The only limitation then involves the interface
thickness relative to the radius of curvature of the dendrites. In
this work we set $\varepsilon=h$ which falls into this range.

The choice of the initial tip radius, $R_0$, for a steady\--state
is not arbitrary. One needs to take into account the latent heat
released at the interface. By choosing a sufficiently large
distance between the interface and the boundary, the latent heat
released at the interface diffuses to the liquid and the effect of
the boundary becomes minimal. In order to guarantee enough
distance to the boundary, the tip radius, $R_0$, should be at
least $20$ times smaller than the diffusion length $D/v_n$. In our
calculation, we choose the tip radius to be $R_0=20{h}$ for the
choice of $N=500$ and $R_{0}=14{h}$ for $N=400$ where $h$ is the
uniform grid size corresponding to the choice of each $N$. Under
these conditions, the diffusion length is large enough and
satisfies the standard theoretical conditions for dendritic
growth~\cite{ProvGDCLKGA1999}.

\section{Parallelization and data distribution}
The numerical simulation of the equations (\ref{Phase1})
and~(\ref{Phase2}) in 3D with any physical choice of parameters is
a difficult task. Of these difficulties, the memory requirements
and the CPU time are the main issues due to the difference in the
length scales. As shown in Table~\ref{tab:Memory}, the demand for
the memory is a delicate issue in that doubling the number of the
grids will increase the computational memory as much as eight
times, and slows down the performance of the code. This makes the
numerical computation of~(\ref{DiscrtePhase1}) and
(\ref{DiscrtePhase2}) with any physically appropriate choice of
grid size impossible on serial computers. Instead, one needs a
parallel architecture in which the work will be distributed to
many computers and the computational job will be shared among the
computers (processors) allowing large scale computation.
\begin{table}[b]
    \begin{center}
           \caption{\textit{Table shows the memory allocation on each processor
           (of 32 processors) when the number of grid points are doubled.}}\smallbreak
       \begin{tabular}{|r|r|r|r}
            \hline
            \em Grid points  & \em Memory(MB)& \em  {~~~Ratio~~~}\\
                \hline
           $32^3$      & $1.45$ & -\\
                \hline
           $64^3$     & $10.56$ & $7.28$\\
                \hline
            $128^3$      &76.90 & $7.28$\\
                \hline
            $216^3$      &611.03 & $7.95$\\
                \hline
        \end{tabular}\label{tab:Memory}
    \end{center}
\end{table}
In this work, we use PETSC's distributed memory architecture~(DA),
whose characteristic feature is that each processor owns its own
local memory, and memory of other processors can not be accessed
directly~\cite{Petsc1997}. The PETSC/DA system requires
communication to inquire or borrow information among the
processors. In particular for the numerical solution of PDE's,
each processor requires its local portion of the information as
well as the points on the boundary of the adjacent processors to
update the right hand side vector. The communication required
among the processors to exchange the components and points along
the border of the adjacent subdomains are managed via the DA
system while the actual data is stored in appropriately sized
local vector objects. Thus, the DA objects only contain the
parallel layout and communication information, and they are not
intended for storing the matrices and vectors.

The communication is necessary but it is very critical that the
parallel code be designed independent of the other processors as
much as possible and the ratio of communication among the
processors should be kept small. Otherwise, a high ratio of
communications among the processors slows down the computation.
Therefore, it is very important that the communication should be
limited to the neighboring processors and should avoid global
communication if possible. Similarly, the distribution of the work
load among the processors is another important issue in parallel
computation. In our work, we keep the number of grid points
proportional to the number of processor so that each processor is
assigned almost the same amount of work load. This enables the
efficient use of the processors and makes the processors to work
in a synchrony. The Table~\ref{tab:Scalability} shows that both
the CPU time and memory allocation are almost halved when the
number of the processors doubled.

The allocation of memory, creation of the parallel matrices and
vectors, and setting up the solver contexts are very time
consuming. Therefore, one would like to use the same initial setup
through out the computation if possible. This is a good approach
especially when the coefficient matrix is independent of time
which is the case in this work. Thus, we can use the same
coefficient matrices as well as the same preconditioners
throughout the calculations.

Parallel solutions of~(\ref{DiscrtePhase1}) and
~(\ref{DiscrtePhase2}) are done via the linear solver of
PETSC~\cite{Petsc1997} in which we use CG iterative method with
Jacobi Preconditioning.
   \begin{table}[tbc]
    \begin{center}
        \caption
        {\textit{The performance of the algorithm is tested for fixed grid (N=128):
        Table shows the wall\--clock time, number of flops and
        the rate of scalability of the algorithm on different number of
        processors.}}\smallbreak
       \begin{tabular}{|r|r|r|r|r|}
            \hline
            \em Number of PE & \em MFlops/s & \em Memory (MB) & \em Wall Clock (s)  & \em {~~Ratio~~}\\
                \hline
            4       &50  &492.8 &1165.78 &-\\
                \hline
            8       &52  &255.5 &574.58  &2.00\\
                \hline
            16      &54  &136.5 &314.15  &3.71\\
                \hline
            32      &55  &76.9 &168.87  &6.90\\
                \hline
            64      &53  &47.0 &88.59   &13.15\\
                \hline
        \end{tabular}\label{tab:Scalability}
    \end{center}
\end{table}
\subsection{Memory allocation and scalability of the algorithm} In
the numerical solution of~(\ref{DiscrtePhase1})
and~(\ref{DiscrtePhase2}), we allocate memory for six parallel and
one sequential global vectors of the size $N^3$, and two parallel
global matrices of the size $N^3\times{N^{3}}$. Together with the
creation of the DA system, local vectors and matrices, the memory
requirement becomes huge for larger $N$. We verify this in
Table~\ref{tab:Memory} by varying the number of grid points. In
fact, it shows that as $N$ is doubled, the memory on each
processor is increased approximately eight times. Thus as N
increases, the demand for the memory gets so large that even for
supercomputers such as Lemieux, there are considerable limitations
of the number when grid points are larger than $N=600$. As
Table~\ref{tab:Memory} indicates, the memory required for a fully
three dimensional computation of phase field model is enormous if
one wants to use a reasonable number of grid points in the
simulation. This necessitates the use of not only high performance
computers but also computers which can accommodate the memory
needed. One way to over come this difficulty is to use parallel
architecture which is the key in our work in handling the memory
deficiency.
   \begin{table}[bh]
    \begin{center}\caption
    {\textit{A sufficient grid size for accurate calculation of interphase growth velocities for SCN is examined.
        The velocities for $u_{\infty}=0.01$ are shown at $t_{final}=16$ sec from the initial stage.}}
      \smallbreak
        \begin{tabular}{|r|r|}
         \hline
            \em \text{Grid Number}  & \em Velocity (cm/sec) \\
            \hline
            $200$       &0.002100\\
                \hline
            $300$       &0.000363\\
                \hline
            $400$       &0.000310\\
                \hline
            $500$       &0.000327\\
                \hline
            $600$       &0.000342\\
                \hline
            $700$       &0.000357\\
                \hline
        \end{tabular}\label{tab:GridConverge}
    \end{center}
\end{table}
The scalability of the code is a measure by which one can test
whether the processors are efficiently used during the
computation. For this we fix the number of the grid points at
$N=128$ and set $T_{final}=300$. By doubling the number of the
processors each time we calculate the corresponding wall\--clock
time for the same job. As shown in the Table~\ref{tab:Scalability}
the wall\--clock time almost doubles when we halved the number of
the processors. This is an indication that the code scales well
with the number of the processors.

As it is apparent from above analysis, memory allocation is still
a delicate issue which influence the choice of grid size N very
much. In our earlier work~\cite{AltunCag}, we studied the grid
convergence in $2D$ by comparing the growth velocities for
different choice of grid points ranging from $200$ to $700$ when
all other parameters kept fixed. Table~\ref{tab:GridConverge}
shows the corresponding velocities for the grid points from $200$
to $700$ when all other conditions are identical for the
undercooling value $0.001$. As seen in
Table~\ref{tab:GridConverge}, the interface growth velocity does
not differ much when we vary the number of grid points, $N$, from
$400$ to $700$ indicating that a choice of $N=500$ will be enough
to study dendritic growth. In this work, we use $N=500$ in the
calculation of undercooling vs. growth velocity linearity
relation. For other calculation such as the study of anisotropy
and kinetic coefficient, we set $N=400$. Also, the time step
$\Delta{t}$ is chosen to be $5\times{10^{-3}}$ throughout the
computation.
\section{Results and conclusions}
In this paper we have performed parallel computations in
three-dimensional space with the specific parameters and boundary
conditions which are used in the microgravity experimental setup
for SCN.

The microgravity experiments exclude most of the convection
effects so that computations involving the physics described by
equations~(\ref{Phase1})-(\ref{Phase2}) or
~(\ref{Sharp1})-(\ref{Sharp3}) can be tested against the
experiment. Prior to these experiments, theoretical results and
computer calculations were awkward in that theory without
convection was tested against experiment (on Earth) with
convection. Thus the interpretation of agreement was ambiguous,
leaving open the possibility of inaccurate computations on
inadequate modeling of experimental setup.

In order to address the questions raised in (A) and (B) of the
introduction, we have considered eight different undercooling
values, $u_{\infty}$, from the microgravity experiments for
SCN~\cite{GlickKossWinsa}. During the simulation we set $R_0=20$,
$N=500$ and $T_{final}=2000$ and compute the average growth
velocity for each undercooling value. The computational and
experimental growth velocities for each undercooling are given in
the Table~\ref{tab:UndIdgeVel}.
    \begin{table}[t]
    \begin{center}
         \caption{\textit{Table shows the computational and experimental interface growth
         velocity for several SCN undercooling values in terms of (cm/sec). First two columns
         contain SCN undercooling values and the corresponding growth velocities from
         space shuttle experiments, respectively.
         The computational results from parallel computing and 3D computation under the
         rotational
         symmetry are given in third and fourth
         columns.}}\smallbreak
         \begin{tabular}{|c|c|c|c|}
          \hline
               \em {~~~~~~\small{$u_\infty$}~~~~~~} &\em Microgravity Vel.
               &\em Rot.Symmet.Vel. &\em Parallel Vel.\\
          \hline
            0.04370       &0.016980 &0.001770 &0.001840\\
               \hline
            0.03380       &0.008720 &0.001486 &0.001480\\
                \hline
            0.02650       &0.004620 &0.001273 &0.001370\\
                \hline
            0.02050       &0.002328 &0.001066 &0.001068\\
                \hline
            0.01610       &0.001417 &0.000922 &0.000902\\
                \hline
            0.01260       &0.000840 &0.000784 &0.000756\\
                \hline
            0.01000       &0.000500 &0.000681 &0.000626\\
                \hline
            0.00790       &0.000343 &0.000590 &0.000456\\
                \hline
        \end{tabular}\label{tab:UndIdgeVel}
    \end{center}
\end{table}

\begin{figure}[hb]
\smallbreak
\begin{center}
\includegraphics*[width=2.7in, height=2.7in]{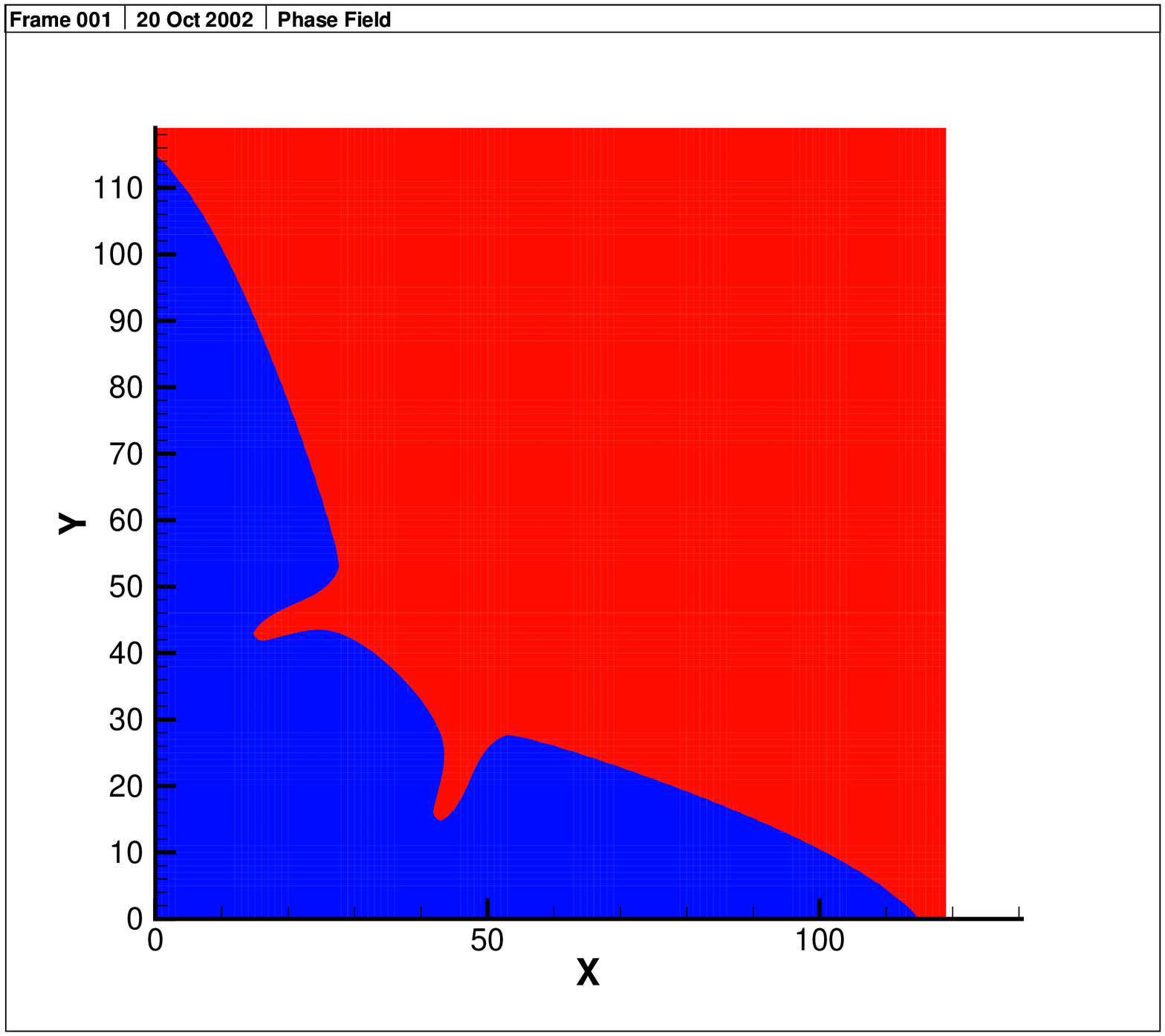}
\includegraphics*[width=2.7in, height=2.7in]{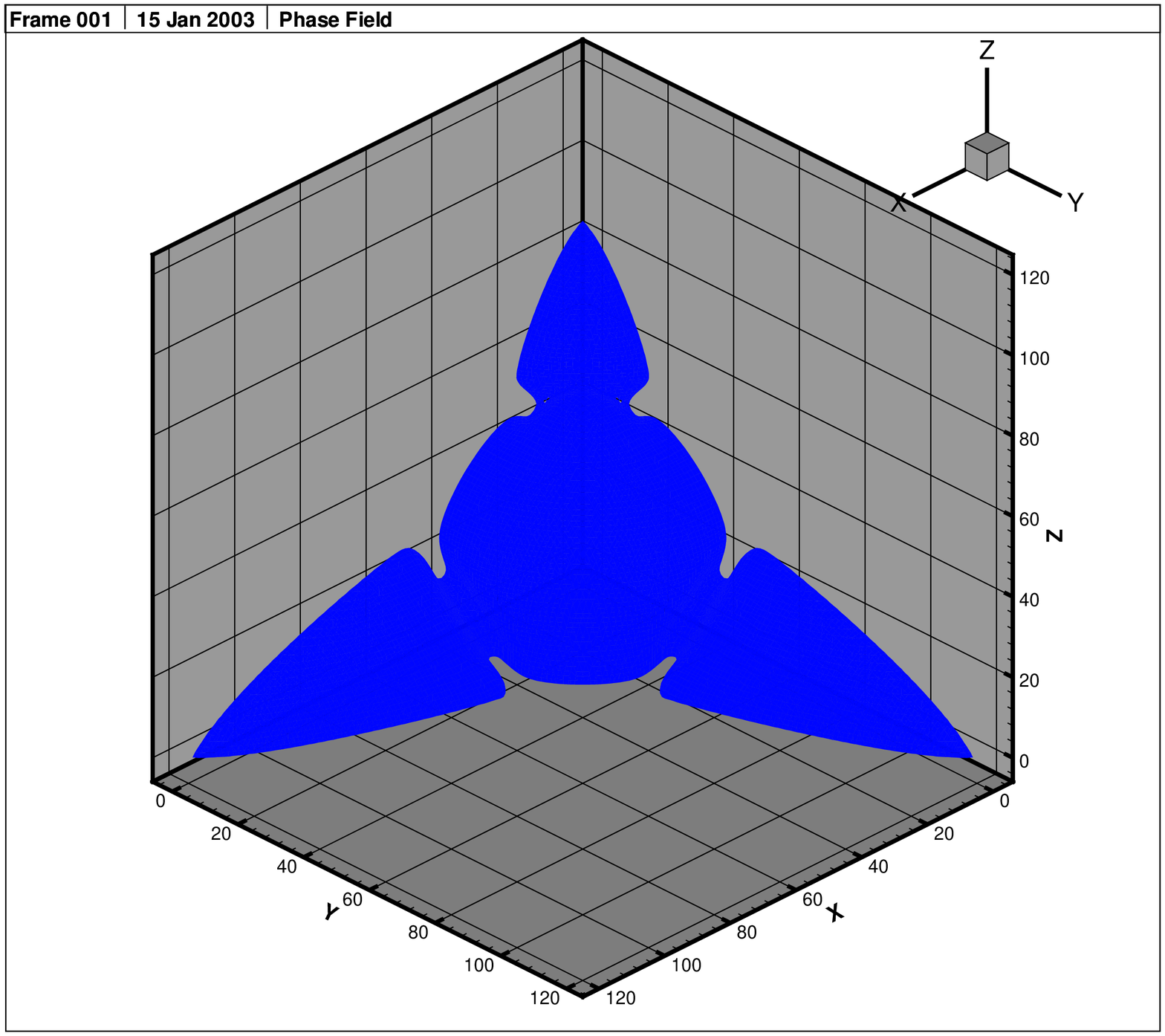}
\end{center}
\caption{\textit{SCN dendrite at~$u_\infty=0.0265$,~ after $215$
seconds: (a) Projection of 3D dendrite on $xy$ plane; (b) 3D SCN
dendrite in the first octant.}} \label{Fig:3D-proj}
\end{figure}
\begin{figure}[tb]
\begin{center}
\includegraphics*[width=2.5in, height=3.5in]{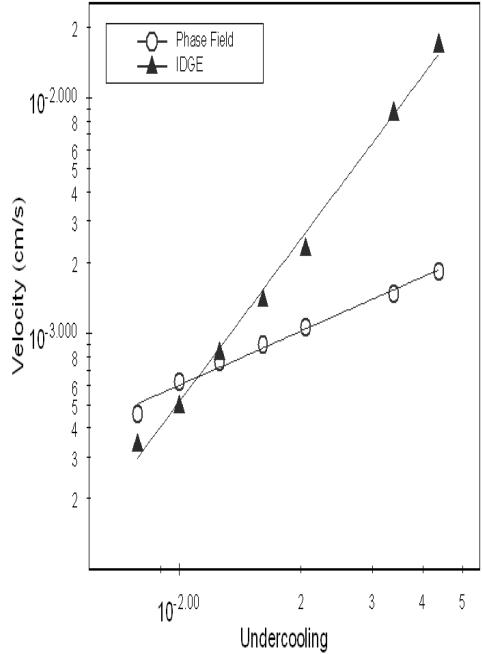}
\end{center}
\caption{\textit{Figure confirms the theoretical result that there
is a linear relation between the undercooling and computational
growth velocity~($u_\infty=0.0265$).}} \label{Fig:3DLinearVel}
\end{figure}

The results in Table~\ref{tab:UndIdgeVel} and corresponding
Fig~\ref{Fig:3DLinearVel} show that computational velocity is
consistent with other 3D phase field computations (see
e.g.~\cite{AltunCag}). The overall results are close to the
experiments, particularly for undercooling temperatures that are
neither very small nor very large. In particular, since
solidification is a complicated process, it is likely that many
other physical effects play a role in determining the growth
velocity at the tip of the dendrite ~\cite{Chalmers}. The
equations~(\ref{Sharp1})-(\ref{Sharp3}) or
equivalently~(\ref{Phase1}) and (\ref{Phase2}) incorporate all of
the physics that have generally been used to study these problems.
Furthermore the numerical schemes are also known to be reliable
through various checks. Hence, it appears that the difference
between our computations and microgravity experiments can be
attributed to additional physics that is not part of the standard
models such as (\ref{Sharp1}) and (\ref{Sharp2}). For the
intermediate values of, such as $0.0126$, the difference is
negligible. Thus, it appears that the model
(equations~(\ref{Phase1}) and~(\ref{Phase2})) includes the key
physical components necessary to describe the solidification
process within this undercooling regime. In particular,
undercoolings that are at the extremes of experimental range, it
is quite likely that simplest physical description given by~
(\ref{Sharp1}), (\ref{Sharp2}), neglects physical factors that are
significant in terms of the growth velocity. At the low end, this
might include, for example, adsorption. At the high end of
undercooling, which generates the higher velocities, the motion of
the dendrite is likely to produce some convective distribution of
heat that would differ from pure heat diffusion in the liquid.
This may be one of the source of randomness or noise that leads to
extensive sidebranching which would lead to additional corrections
to the velocity. At the present there are no coherent methods to
incorporate noise into the phase field (or sharp interface)
equations. In the absence of experimental data on interface noise,
the use of noise in computations would involve at least two ad hoc
parameters (amplitude and frequency), so that any resulting
agreement with the experimental data would not be very meaningful.

The model includes several features such as surface tension and
kinetic undercooling. The importance of surface tension
(manifested in the capillarity length, $d_{0}$) has been noted in
studies of linear stability~\cite{Ockendon1980,MSekerka} and
computations. However, the kinetic undercooling, $\alpha$, is
often neglected in computations and theoretical studies. In fact,
while the phase field equations naturally incorporate this
material parameter, some calculations have used a limit in which
$\alpha$ approaches zero, rather than its true value, for
computational convenience. We perform pairs of calculations for
the undercooling value $0.0205$ in which all parameters were
identical except for $\alpha$ (see Table~\ref{tab:KinVel}).
    \begin{table}[h]
    \begin{center}
    \caption{\textit{Table shows the effect of kinetic coefficient on the growth velocity of interface
    for the undercooling $u_\infty=0.0205$.}}
    \begin{tabular}{|c|c|c|}
    \hline
    \em $\alpha (sec/cm^2)$    &\em Velocity(cm/sec)\\
    \hline
            2.5x$10^6$       &0.00117\\
                \hline
            3.5x$10^6$       &0.00090\\
                \hline
            5.0x$10^6$       &0.00106\\
                \hline
        \end{tabular}\label{tab:KinVel}
        \end{center}
\end{table}
These results indicate that a change in the kinetic coefficient of
28.5\% can result in a growth velocity that is 30\% as shown in
Table \ref{tab:KinVel}. This suggests that $\alpha$ ( as
$\beta:={\alpha}d_{0}$ as the computation of the velocity in
Gibbs\-- Thomson relation~(\ref{Sharp3})) can not be used as a
free parameter, as can $\varepsilon$, the interface thickness.
Moreover, variation in $\alpha$ implies a change in growth
velocity, as does a variation in any of the other parameters such
as  $c_{v}$, $l_{v}$, $K$, $etc.$. In all other computations, we
set $\alpha$ to be $3.5\times{10^6}sec/{cm^2}$ which is
approximated by using SCN microgravity values in~(\ref{Sharp3}).

Our results have some interesting implications for dimensionality,
as we can compare our 3D calculations to 2D calculations and to
rotationally symmetric 3D calculations in which we used
cylindrical coordinates and assumed that the dependence was purely
radial. The experimental pictures indicate that the cylindrical
symmetry of the single-needle crystal breaks down shortly beyond
the tip. Our results, however, indicate that there is relatively
little difference in velocity between the two calculations. The
tip velocity calculations in 3D and 2D, on the other hand, differ
by about a factor of $1.9$ (see Table~\ref{tab:Comp2D3DVel}).
    \begin{table}[b]
    \begin{center}
        \caption{\textit{The table shows the computational interface growth velocities (cm/sec) in $2D$
        and 3D (parallel) for different undercooling
        values.}}\smallbreak
        \begin{tabular}{|c|c|c|}
            \hline
            \em Undercooling        &\em 2D\--Velocity   & \em 3D\--Velocity\\
                \hline
            0.01         &0.00033     &0.00063\\
                \hline
            0.0161       &0.00050     &0.00090\\
                \hline
            0.0265       &0.00066     &0.00137\\
                \hline
        \end{tabular}\label{tab:Comp2D3DVel}
    \end{center}
\end{table}
The ratio $1.9$ can be put in perspective by examining the
limiting sharp interface equations (\ref{Sharp1}) and
(\ref{Sharp2}). Physical intuition suggests that the growth of the
interface is limited mainly by the diffusion of the latent heat
manifested in the condition (\ref{Sharp2}). When diffusion is
rapid, the heat equation is approximated by Laplace's equation,
whose radial solutions are of the form $r^d$. The latent heat
condition (\ref{Sharp2}) implies that the normal velocity is
proportional to the gradient, or $dr^{d-1}$. Comparing this term
for $d = 3$ versus $d = 2$, one has a ratio of $3/2 = 1.5$.
Analogously, if we examine the Gibbs\--Thomson relation alone, and
solve (\ref{Sharp3}) for the normal velocity, we see that
dimensionality arises (directly) in terms of $k$, the sum of
principal curvatures, which is $(d - 1)/R_0$ where $R_0$ is the
radius of curvature. Hence this factor would suggest that at least
one of the terms in this expression for the velocity has a
coefficient $d - 1$, suggesting a ratio of $(3 - 1)/(2 - 1) = 2$.
Thus a heuristic examination of the key limiting equations
suggests that the tip velocity in $3D$ should be about $1.5$ to
$2$ times that of the $2D$ system. Of course there are numerous
nonlinearities involved in the equations that could alter this
ratio. Our calculations fall well in the range $1.5$ to $2$,
thereby lending some support to the heuristics above.
\begin{figure}[t]
\begin{center}
\includegraphics*[width=4in, height=2in]{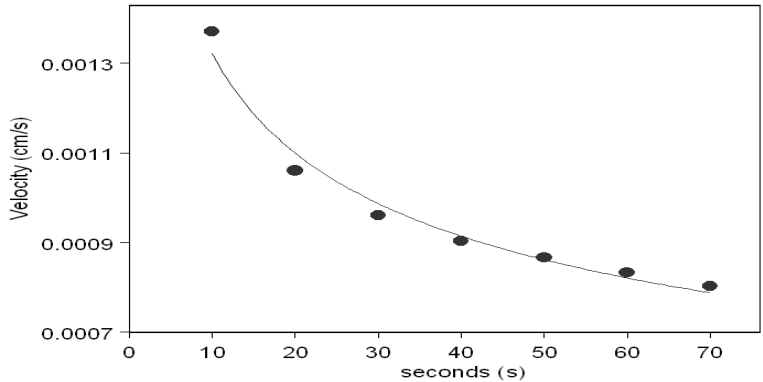}
\end{center}
\caption{\textit{Figure confirms the theoretical result that the
growth velocity approaches a constant value in large
time~($u_\infty=0.0265$).}}\label{Fig:VelLargeTime}
\end{figure}
The fully three dimensional calculations also allow a complete
treatment of the anisotropy as it is manifested in both directions
(see equation~\ref{SurfaceTensionN}). While the immediate area
near the tip of the dendrite appears to be symmetric about the
direction of growth, the photographs of experiments show that
there is significant asymmetry a short distance away from the tip.
Consequently, there is some question as to the accuracy of
rotationally symmetric computations (reducing the 3D problem to
one which is a 2D computation). However, we find that this
asymmetry influences the tip velocity by only 8-10\%. Nevertheless
this anisotropy can be expected to play a key role in the
development of sidebranching for which the axial symmetry appears
to be significant.

To verify the effect of the surface tension anisotropy on the
interface growth, we use four different anisotropy levels, $0.00$,
~$0.006$, ~$0.009$ ~and ~$0.01$ for the undercooling value
$0.0205$. Corresponding growth velocities are~$9.1 \times 10^{-4}~
cm/sec$,~$1.16\times 10^{-3} cm/sec$, ~$1.20 \times 10^{-3}
cm/sec$ and~$1.07 \times 10^{-3} cm/sec$, respectively. The
influence of the anisotropy on the shape of the interface is more
clear compared to the effects on the growth velocity(see
Fig~\ref{Fig:ContourPlots}). An order of magnitude change in the
anisotropy strength does not change tip velocity significantly
which confirm the experimental result~\cite{GlickSingh1989} as
well as the results from rotational symmetry case we studied
in~\cite{AltunCag}. We also observe that the tip of the interface
becomes sharper in the preferred direction as the strength of the
anisotropy increases.

As indicated in experiments, theory and computational
studies~\cite{KimProvGoldDant1999}, the average growth rate of a
single needle-crystal in 3D approaches a constant value. We
examine this issue by using the undercooling value,
$u_\infty=0.0205$, for SCN. The average growth velocities at
different time steps are calculated with the anisotropy strength
$\delta_{\sigma}=0.01$. The growth velocity approaches a constant
value as $T_{final}$ gets larger (see Fig~\ref{Fig:VelLargeTime})
which confirms previous computational and theoretical
studies~\cite{KimProvGoldDant1999}.

We have performed all of our calculations on the terascale
computing system, Lemieux, at Pittsburgh Super Computing Center.
Lemieux consists of $750$ Compaq Alphaserver $ES45$ nodes and two
separate front end nodes. Each node contains four 1-GHz processors
SMP with $4$ Gbytes of memory.

\end{document}